\documentclass[12pt]{article}

\usepackage{newtxtext,newtxmath}

\usepackage{graphicx}

\usepackage[letterpaper,margin=1in]{geometry}

\renewenvironment{abstract}
	{\quotation}
	{\endquotation}

\date{}

\makeatletter
\renewcommand{\fnum@figure}{\textbf{Figure \thefigure}}
\renewcommand{\fnum@table}{\textbf{Table \thetable}}
\makeatother

\usepackage{scicite}

\usepackage{url}

\def\scititle{
	Humidity-modulated Ostwald Step Rule in $\rm FAPbI_3$ Martensitic Transitions
}
\title{\bfseries \boldmath \scititle}

\author{
	Ping Tuo$^{1\ast}$,
	Ju Li$^{2\ast}$\and
	\small$^{1}$Bakar Institute of Digital Materials for the Planet, University of California, Berkeley, United States.\and\
	\small$^{2}$Department of Nuclear Science and Engineering and Department of Materials Science and Engineering,\\ 
    \small Massachusetts Institute of Technology, Cambridge, United States.\and
	\small$^\ast$Corresponding author. Email: tuoping@berkeley.edu, liju@mit.edu.
}

\begin{document} 

\maketitle

\begin{abstract}  
Experimental observations reveal a dual role of moisture in $\rm FAPbI_3$ phase transitions. During fabrication at high temperatures, humidity improves the quality of the perovskite ($\alpha$ phase) films, whereas at low service temperatures, humidity accelerates degradation of the photoactive $\alpha$ phase into the photo-inactive hexagonal $\delta$ phase. Despite extensive experimental evidence, the microscopic mechanisms by which moisture modulates phase stability and transition pathways remain unknown. 
Here, using large-scale atomistic simulations powered by an ab initio machine-learned potential, we reproduce the experimentally observed increase in the $\delta$-to-$\alpha$ transition temperature under increased humidity. Our simulations reveal a layered metastable intermediate along the Burgers transition pathway that exhibits a strong affinity for water. The competing water uptake of this intermediate and the $\delta$ phase governs the humidity dependence of transition kinetics. 
These findings elucidate the dual role of moisture in perovskite phase transitions and demonstrate how environmental species can stabilize metastable intermediates and reshape transition pathways, providing an environment-dependent perspective on Ostwald's step rule.
\end{abstract}

\noindent
Moisture has long been recognized as a double-edged sword in the phase stability and synthesis of $\rm FAPbI_3$. On the one hand, a sharp decline in optical absorptance has been repeatedly observed when the relative humidity exceeds approximately $\rm 45\%$~\cite{raval2022understanding,yun2018humidity}, indicating moisture-induced degradation of the photoactive perovskite ($\alpha$) phase. On the other hand, moisture exposure has been shown to accelerate the $\delta$-to-$\alpha$ phase transition and improve the perovskite film quality during fabrication~\cite{lin2022vapor,jheng2026realizing,liu2022moisture,ma2025crystallization,cao2024interactions}. Despite extensive experimental investigations, the microscopic mechanisms underlying these effects remain poorly understood. Elucidating how moisture influences the relative phase stability and transition pathways of $\rm FAPbI_3$ could therefore not only guide the development of more stable perovskite photovoltaics but also provide broader insights into the role of humidity in solid-state phase transitions.

Here, we unravel the humidity-dependent phase behavior of $\rm FAPbI_3$ through large-scale atomistic simulations powered by a machine-learned interatomic potential with ab initio accuracy. By calculating and comparing the free energies of $\alpha$ and $\delta$ phases across a broad range of temperatures and relative humidities, we establish a thermodynamic phase diagram that reveals an increasing $\delta\rightarrow\alpha$ transition temperature under higher humidity, in good agreement with experimental observations. More importantly, we identify a layered metastable intermediate along the transition path that exhibits a pronounced affinity for water uptake. The competition of humidity-dependent water uptake between this intermediate and the $\delta$ phase modulates the transition kinetics and controls the speed of $\alpha\rightarrow\delta$ degradation. These findings provide a microscopic understanding of how moisture affects solid-solid phase transitions by modulating the formation and stability of metastable intermediates, providing new insights into the role of environmental species in Ostwald's step rule.

\section*{$\alpha-\delta$ transition under humidity}

\begin{figure}[htbp]
    \centering
    \includegraphics[width=1\linewidth]{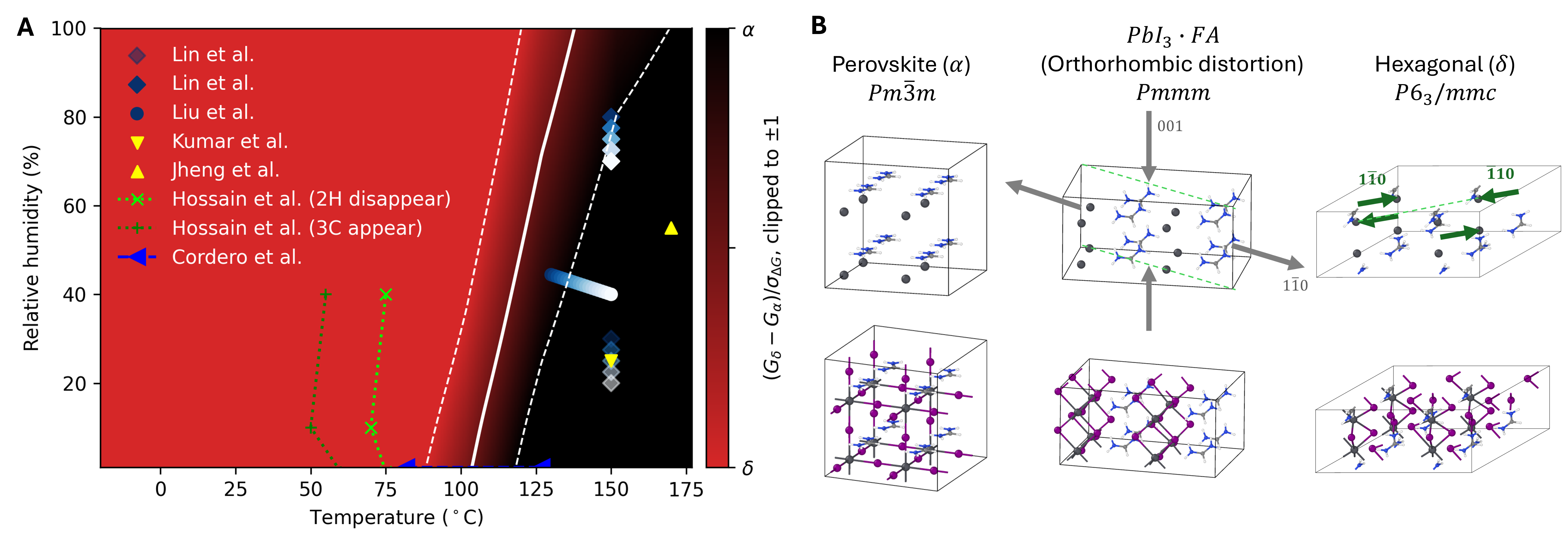}
    \caption{\textbf{Phase diagram of $\rm FAPbI_3$}. (\textbf{A}) Temperature-humidity phase diagram of $\rm FAPbI_3$ and comparison with experimental observations. The colour map represents the normalized Gibbs free-energy difference between the $\alpha$ and $\delta$ phases as a function of temperature and relative humidity (RH). The solid white curve marks the predicted phase boundary, where the two phases have equal free energies, while the dashed white curves delimit the uncertainty in the predicted boundary. Experimental observations from six studies are overlaid for comparison. The experimental results are taken from refs.~\cite{cordero2020influence,cordero2021structural,lin2022vapor,liu2022reversible,kumar2024superior,jheng2026realizing,hossain2026balancing}. (\textbf{B}) Illustration of the Burgers path of the $\rm FAPbI_3$ $\alpha\leftrightarrow\delta$ transition. }
    \label{fig:burgers}
\end{figure}

To predict the phase transition temperature under different humidity, we first use thermal integration to calculate the free energies of $\alpha$ and $\delta$ phases at different temperatures, then run grand canonical Monte Carlo of water insertion to establish a relation between the grand canonical free energy $\Omega$ and relative humidity ($\rm RH$). The $\delta$ phase equilibrates with high water uptake at increased $\rm RH$, while the $\alpha$ phase shows almost zero water uptake.
As a result, the transition temperature rises with increased $\rm RH$ (Figure~\ref{fig:burgers}a). Here, we used the relationship between $\Omega$ and $\rm RH$ determined at $\rm 76.85~^\circ C$ ($\rm 350~K$) as an approximation for higher temperatures.

We compare the predicted phase boundary with experimental observations from six studies in Figure~\ref{fig:burgers}A, finding excellent agreement overall. 
Cordero et al.~\cite{cordero2020influence,cordero2021structural} reported thermal hysteresis in the $\alpha \leftrightarrow \delta$ transition, spanning $\rm 77~^\circ C$ upon cooling and $\rm 127~^\circ C$ upon heating. They further demonstrated increased transition temperature under trace humidity.
Lin et al.~\cite{lin2022vapor} investigated humidity treatments at $\rm RH$ values $\rm 30\%$ and $\rm 80\%$, followed by annealing at $\rm 150~^\circ C$, and reported improved photovoltaic performance under the higher-humidity treatment owing to reduced side-products. Liu et al.~\cite{liu2022reversible} demonstrated reversible phase transitions through ``humidity aging-heat healing'' cycles, in which the $\alpha$ phase transformed into the $\delta$ phase at $\rm RH \approx 60\%$ and subsequently reverted to the $\alpha$ phase upon annealing at $\rm 150~^\circ$C. Kumar et al.~\cite{kumar2024superior} achieved a stable $\alpha$ phase by annealing $\delta$-phase films at $\rm 150~^\circ$C under $\rm RH$ $\rm 25\%$. Similarly, Jheng et al.~\cite{jheng2026realizing} obtained high-quality, phase-pure $\alpha$-$\rm FAPbI_3$ films with enhanced power conversion efficiency through annealing at $\rm 170~^\circ C$ under $\rm RH$ as high as $\rm 55\%$ and $\rm 99\%$. 
During film deposition, Hossain et al.~\cite{hossain2026balancing} observed the emergence of the kinetically favored $\alpha$ phase at approximately $\rm 60~^\circ$C and the disappearance of $\delta$-phase at approximately $\rm 75~^\circ C$ across a range of humidity levels.

\section*{Kinetics modulation by water uptake of an intermediate phase}

$\rm FAPbI_3$ perovskite assumes a body-centered cubic (bcc) lattice structure, with $\rm Pb$ and $\rm FA$ occupying the edge and body-center, respectively (Figure~\ref{fig:burgers}b). A classic mechanism of the transition between bcc and hexagonal lattices is the Burgers path. This pathway involves two steps: the first step is an orthorhombic lattice deformation, where the lattice is stretched in the [110] direction and compressed in the [001] direction, and the second step is a cooperative atom shuffling in the [110] direction with atoms in the adjacent planes shifting in opposite directions. The first step of lattice deformation transforms the geometry of the $\{110\}$ planes of the bcc lattice into the basal planes of the hcp lattice, with mild atomic displacements. The second step shifts the lattice from orthorhombic to hexagonal.
For $\rm FAPbI_3$, the Burgers path is accompanied by breaking and reforming of $\rm Pb-I$ bonds, as illustrated in Figure~\ref{fig:burgers}B. 


The crystallographic anisotropy associated with the Burgers path is reflected in two complementary experimental observations. First, moisture-induced degradation of the $\alpha$ phase exhibits a pronounced facet dependence, with the \{100\} facet being more susceptible to degradation than the \{111\} facet~\cite{ma2023unveiling}. Second, termination of the $\delta$ phase along its \{110\} planes by introducing two-dimensional perovskitoid materials into the perovskite matrix has been demonstrated as an effective strategy for stabilizing FA-based perovskite solar cells~\cite{xing2026mechanistic} by suppressing the Burgers path.

\begin{figure}[htbp]
    \centering
    \includegraphics[width=1\linewidth]{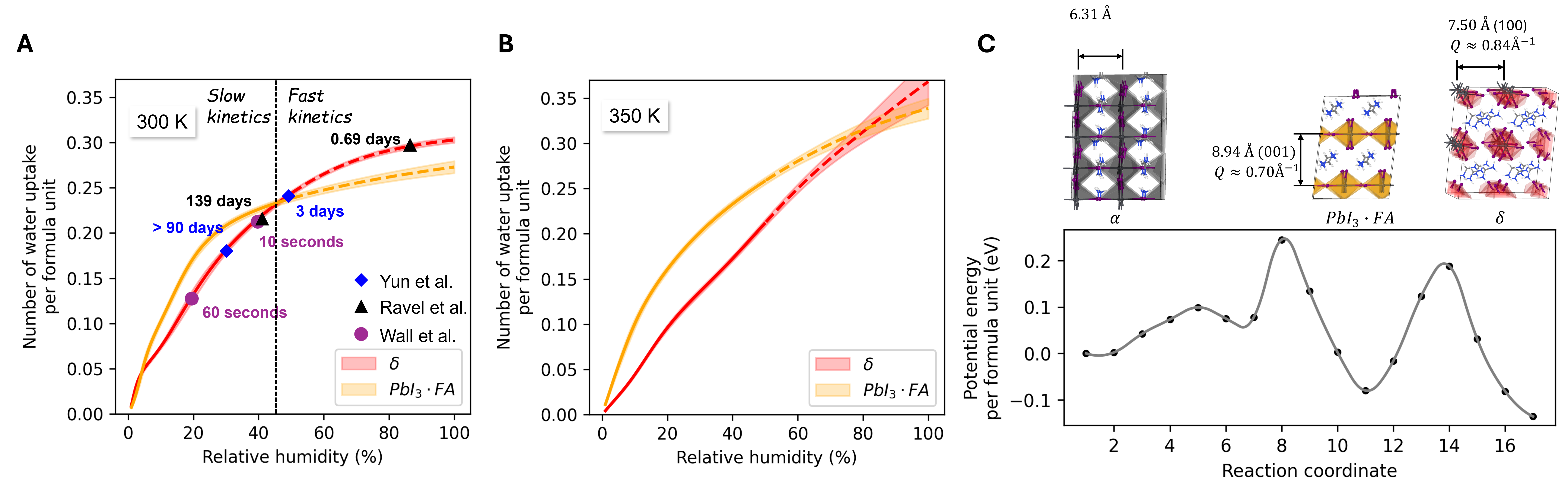}
    \caption{\textbf{Water uptake of the $\delta$ and intermediate $\rm PbI_3\cdot FA$ phases.} (\textbf{A}) Water uptake per formula unit in the $\delta$ and intermediate $\rm PbI_3\cdot FA$ phases obtained at $\rm 26.85~^\circ C$ ($\rm 300~K$). The dashed portion of the line is extrapolated from sampling results obtained at lower $\rm RH$ values. The $\alpha$ phase exhibits negligible water uptake in all grand canonical Monte Carlo (GCMC) simulations and is therefore omitted. Experimentally reported $\alpha\rightarrow\delta$ transition times at different humidity levels are marked for comparison~\cite{yun2018humidity,wall2026impact,raval2022understanding}. (\textbf{B}) Water uptake per formula unit in the $\delta$ and intermediate $\rm PbI_3\cdot FA$ phases obtained at $\rm 76.85~^\circ C$ ($\rm 350~K$). The $\alpha$ phase exhibits negligible water uptake and is omitted. (\textbf{C}) Burgers transition pathway connecting the $\alpha$ and $\delta$ phases via the intermediate $\rm PbI_3\cdot FA$ phase, optimized using the nudged elastic band (NEB) method. The first diffraction peak positions and corresponding interplanar spacings are annotated for each phase.}
    \label{fig:kinetics}
\end{figure}

More intriguingly, the first step along the Burgers pathway gives rise to a layered metastable intermediate phase ($\rm PbI_3\cdot FA$) with a pronounced affinity for water uptake. As shown in Figure~\ref{fig:kinetics}A, at room temperature, water uptake of the intermediate increases sharply with increasing $\rm RH$, whereas that of the $\delta$ phase increases more gradually but eventually surpasses that of the intermediate at $\rm RH \gtrsim 45\%$.  
At the elevated temperature of $\rm 76.28~^\circ C$, however, the intermediate exhibits greater water uptake than the $\delta$ phase across a much wider $\rm RH$ range (Figure~\ref{fig:kinetics}B). This temperature-dependent contrast provides a possible explanation for the dual effects of humidity on phase transitions observed experimentally.

First, multiple studies have reported a drastic drop in the lifetime of $\alpha$-$\rm FAPbI_3$ from several months to only a few days, once $\rm RH$ levels exceed approximately $\rm 45\%$ (Figure~\ref{fig:kinetics}A)~\cite{wang2022phase,yun2018humidity,wall2026impact,raval2022understanding}. This can be explained by the competition for water uptake between the intermediate and $\delta$ phases shown in Figure~\ref{fig:kinetics}A. At $\rm RH<\rm 45\%$, the intermediate phase adsorbs more water than the $\delta$ phase. Therefore, the second step of the Burgers path from the intermediate phase to the $\delta$ phase involves expulsion of the water interstitials settled in the lattice. This creates an extra barrier and stalls the formation of the $\delta$ phase, resulting in a longer $\alpha$ phase lifetime. When $\rm RH>45\%$, on the other hand, water uptake of the $\delta$ phase surpasses that of the intermediate phase, and the $\alpha\rightarrow\delta$ phase degradation is accelerated drastically. 
Second, a trace XRD peak at $Q \approx 0.7~ \rm \mathring{A}^{-1}$ has been repeatedly observed at medium $\rm RH$ but disappears at higher RH~\cite{liu2022reversible,wang2021moisture,wang2023stable,ma2016stable}. The corresponding interplanar spacing of approximately $\rm 8.95~\mathring{A}$ is consistent with the \{001\} planes of the metastable intermediate phase, suggesting its transient formation during humidity-driven phase transition (Figure~\ref{fig:kinetics}C). 

Conversely, during fabrication, humidity has been widely employed to help accelerate the $\alpha$ phase formation and improve the quality of the perovskite film ~\cite{lin2022vapor,jheng2026realizing,liu2022moisture,ma2025crystallization,cao2024interactions}. 
At the elevated fabrication temperatures ($\gtrsim \rm 150~^\circ C$), the water uptake is always higher in the intermediate phase (Figure~\ref{fig:kinetics}B), promoting the second step of the Burgers path from $\delta$ to the intermediate phase and reducing the leftover $\delta$ phase impurities in the perovskite film. Therefore, even $\rm RH$ as high as $\rm 99\%$ has been reported to improve the lifetime of the produced perovskite films~\cite{jheng2026realizing}.  

Together, these findings support the dual effect of humidity, where humidity can either accelerate $\alpha\rightarrow\delta$ degradation or assist $\delta\rightarrow\alpha$ phase conversion and reduce $\delta$ phase impurities in the fabricated perovskite film, depending on temperature and the relative water affinities of the competing intermediate and $\delta$ phases.


The implications of this mechanism extend beyond perovskites. Layered polymorphs are widespread among ionic solids, including halides, oxides, sulfides, and hydrated minerals, where environmental species such as water can intercalate into and stabilize layered structures. We therefore propose that humidity-mediated stabilization of metastable intermediates may represent a broader mechanism governing phase transitions in ionic solids. Our findings suggest an environment-mediated extension of Ostwald's step rule, in which external species actively reshape the free-energy landscape, stabilize otherwise transient intermediates, and modulate transition dynamics.


\clearpage 

%
\bibliography{science_template} 

@article{ma2025crystallization,
  title={Crystallization-activated moisture barrier for high-tolerance manufacturing of perovskite solar cells},
  author={Ma, Yabin and Luo, Chao and Zhan, Changling and Wang, Keli and He, Jiandong and Gao, Peng and Bi, Zhuye and Ma, Yingzhuang and Zhao, Qing},
  journal={Science Advances},
  volume={11},
  number={43},
  pages={eady5703},
  year={2025},
  publisher={American Association for the Advancement of Science}
}

@article{liu2022moisture,
  title={Moisture-triggered fast crystallization enables efficient and stable perovskite solar cells},
  author={Liu, Kaikai and Luo, Yujie and Jin, Yongbin and Liu, Tianxiao and Liang, Yuming and Yang, Liu and Song, Peiquan and Liu, Zhiyong and Tian, Chengbo and Xie, Liqiang and others},
  journal={Nature Communications},
  volume={13},
  number={1},
  pages={4891},
  year={2022},
  publisher={Nature Publishing Group UK London}
}

@article{raval2022understanding,
  title={Understanding instability in formamidinium lead halide perovskites: kinetics of transformative reactions at grain and subgrain boundaries},
  author={Raval, Parth and Kennard, Rhiannon M and Vasileiadou, Eugenia S and Dahlman, Clayton J and Spanopoulos, Ioannis and Chabinyc, Michael L and Kanatzidis, Mercouri and Manjunatha Reddy, GN},
  journal={ACS Energy Letters},
  volume={7},
  number={4},
  pages={1534--1543},
  year={2022},
  publisher={ACS Publications}
}

@article{yun2018humidity,
  title={Humidity-induced degradation via grain boundaries of {HC(NH$_2$)$_2$PbI$_3$} planar perovskite solar cells},
  author={Yun, Jae Sung and Kim, Jincheol and Young, Trevor and Patterson, Robert J and Kim, Dohyung and Seidel, Jan and Lim, Sean and Green, Martin A and Huang, Shujuan and Ho-Baillie, Anita},
  journal={Advanced Functional Materials},
  volume={28},
  number={11},
  pages={1705363},
  year={2018},
  publisher={Wiley Online Library}
}

@article{lin2022vapor,
  title={Vapor deposited pure $\alpha$-{FAPbI$_3$} perovskite solar cell via moisture-induced phase transition strategy},
  author={Lin, Dongxu and Gao, Yujia and Zhang, Tiankai and Zhan, Zhenye and Pang, Nana and Wu, Zongwang and Chen, Ke and Shi, Tingting and Pan, Zhenqiang and Liu, Pengyi and others},
  journal={Advanced functional materials},
  volume={32},
  number={48},
  pages={2208392},
  year={2022},
  publisher={Wiley Online Library}
}

@article{jheng2026realizing,
  title={Realizing High-Performance Vacuum-Deposited Inverted $\alpha$-{FAPbI$_3$} Perovskite Solar Cells through Saturated-Humidity Annealing},
  author={Jheng, Yun-Sheng and Chen, Cheng-Yueh and Jan, Pei-En and Chen, Hung-Ming and Lin, Hao-Cheng and Tsai, Ping-Hsun and Li, Chia-Feng and Huang, Yu-Ching and Truong, Minh Anh and Wakamiya, Atsushi and others},
  journal={ACS Applied Materials \& Interfaces},
  volume={18},
  number={12},
  pages={18049},
  year={2026}
}

@article{xing2026mechanistic,
  title={Mechanistic insight into $\alpha$-to-$\delta$ phase transition and stabilizing $\alpha$-phase {FAPbI$_3$} via regulating $\delta$-phase orientation},
  author={Xing, Chuwu and Li, Yongzhe and Bao, Qinhui and He, Miao and Lu, Zhiqiang and Zhang, Yibin and Liu, Zhike and Zhang, Tianjin and Wang, Duofa},
  journal={Nature Communications},
  year={2026},
  publisher={Nature Publishing Group}
}

@article{cao2024interactions,
  title={Interactions between {H$_2$O} and lead halide perovskites: Recent progress and applications},
  author={Cao, Jingjing and Zhang, Xiaoyu and Miao, Yanfeng and Li, Wen and Zeng, Xiankan and Yang, Shiyu and Yan, Cheng and Lu, Jun and Yang, Weiqing},
  journal={Matter},
  volume={7},
  number={11},
  pages={3728--3755},
  year={2024},
  publisher={Elsevier}
}

@article{ma2023unveiling,
  title={Unveiling facet-dependent degradation and facet engineering for stable perovskite solar cells},
  author={Ma, Chunqing and Eickemeyer, Felix T and Lee, Sun-Ho and Kang, Dong-Ho and Kwon, Seok Joon and Gr{\"a}tzel, Michael and Park, Nam-Gyu},
  journal={Science},
  volume={379},
  number={6628},
  pages={173--178},
  year={2023},
  publisher={American Association for the Advancement of Science}
}

@article{liu2022reversible,
  title={Reversible Phase Transition for Durable Formamidinium-Dominated Perovskite Photovoltaics},
  author={Liu, Huifen and Li, Nengxu and Chen, Zehua and Tao, Shuxia and Li, Chunlei and Jiang, Lang and Niu, Xiuxiu and Chen, Qi and Wang, Feng and Zhang, Yu and others},
  journal={Advanced Materials},
  volume={34},
  number={39},
  pages={2204458},
  year={2022},
  publisher={Wiley Online Library}
}

@article{kumar2024superior,
  title={Superior stabilized $\alpha$-{FAPbI$_3$} perovskite solar cells with efficiency exceeding 24\%},
  author={Kumar, Anjan and Sharma, Pawan and Ved, Amit and Abd Hamid, Junainah and Mohammed, Adil Ismael and Singh, Ashish and Kaushik, Vikas and others},
  journal={Organic Electronics},
  volume={135},
  pages={107143},
  year={2024},
  publisher={Elsevier}
}

@article{cordero2021structural,
  title={Structural transitions and stability of {FAPbI$_3$} and MAPbI3: The role of interstitial water},
  author={Cordero, Francesco and Craciun, Floriana and Paoletti, Anna Maria and Zanotti, Gloria},
  journal={Nanomaterials},
  volume={11},
  number={6},
  pages={1610},
  year={2021},
  publisher={MDPI}
}

@article{cordero2020influence,
  title={Influence of temperature, pressure, and humidity on the stabilities and transition kinetics of the various polymorphs of {FAPbI$_3$}},
  author={Cordero, Francesco and Craciun, Floriana and Trequattrini, Francesco and Generosi, Amanda and Paci, Barbara and Paoletti, Anna Maria and Zanotti, Gloria},
  journal={The Journal of Physical Chemistry C},
  volume={124},
  number={42},
  pages={22972--22980},
  year={2020},
  publisher={ACS Publications}
}

@article{hossain2026balancing,
  title={Balancing moisture and oxygen can match the crystallization dynamics of inert halide perovskite processing},
  author={Hossain, Maimur and Dolan, Connor J and Oberholtz, Eric and Kamiyama, Darya and Palmer, Jack R and Marchezi, Paulo E and Kodalle, Tim and Sutter-Fella, Carolin M and Fenning, David P},
  journal={EES Solar},
  volume={2},
  number={3},
  pages={715--723},
  year={2026},
  publisher={The Royal Society of Chemistry}
}

@article{wang2021moisture,
  title={Moisture-stable {FAPbI$_3$} perovskite achieved by atomic structure negotiation},
  author={Wang, Ryan Taoran and Xu, Alex Fan and Li, Wuqi and Li, Yuning and Xu, Gu},
  journal={The Journal of Physical Chemistry Letters},
  volume={12},
  number={22},
  pages={5332--5338},
  year={2021},
  publisher={ACS Publications}
}

@article{wang2023stable,
  title={Stable {FAPbI$_3$} hydrate structure by kinetics negotiation for solar cells},
  author={Wang, Ryan Taoran and Zhang, Yu and Wu, Xiaoxue and Zhang, Weiwei and Chi, Longxing and Xu, Fan},
  journal={Sustainable Energy \& Fuels},
  volume={7},
  number={8},
  pages={1974--1980},
  year={2023},
  publisher={The Royal Society of Chemistry}
}

@article{ma2016stable,
  title={Stable $\alpha$/$\delta$ phase junction of formamidinium lead iodide perovskites for enhanced near-infrared emission},
  author={Ma, Fusheng and Li, Jiangwei and Li, Wenzhe and Lin, Na and Wang, Liduo and Qiao, Juan},
  journal={Chemical science},
  volume={8},
  number={1},
  pages={800--805},
  year={2016},
  publisher={The Royal Society of Chemistry}
}

@article{wang2022phase,
  title={Phase-Transition-Cycle-Induced Recrystallization of {FAPbI$_3$} Film in An Open Environment Toward Excellent Photodetectors with High Reproducibility},
  author={Wang, Meng and Cao, Fengren and Meng, Linxing and Wang, Min and Li, Liang},
  journal={Advanced Science},
  volume={9},
  number={34},
  pages={2204386},
  year={2022},
  publisher={Wiley Online Library}
}

@article{wall2026impact,
  title={Impact of Processing Environment on Anti-Solvent Free {FAPbI$_3$} Films and Solar Cells},
  author={Wall, Elizabeth M and Lin, Yu-Kuan and Bawendi, Moungi and Burlingame, Quinn C and Loo, Yueh-Lin},
  journal={Advanced Energy Materials},
  volume={16},
  number={7},
  pages={e03828},
  year={2026},
  publisher={Wiley Online Library}
}
\bibliographystyle{sciencemag}

%
%
%
%
%
%


\section*{Acknowledgments}

The authors acknowledge the research computing facilities provided by the Bohrium Cloud Platform and the Savio computational cluster resource provided by the Berkeley Research Computing program at the University of California, Berkeley (supported by the UC Berkeley Chancellor, Vice Chancellor for Research, and Chief Information Officer). P.T. acknowledges valued discussions with Dr. Linfeng Zhang, Dr. Michele Invernizzi, Dr. Haiyang Niu, Dr. Manyi Yang, Dr. Fuzhi Dai, and Dr. Han Wang.

\paragraph*{Funding:}
P.T. acknowledges funding from the BIDMaP Postdoctoral Fellowship.

\newpage


\renewcommand{\thefigure}{S\arabic{figure}}
\renewcommand{\thetable}{S\arabic{table}}
\renewcommand{\theequation}{S\arabic{equation}}
\renewcommand{\thepage}{S\arabic{page}}
\setcounter{figure}{0}
\setcounter{table}{0}
\setcounter{equation}{0}
\setcounter{page}{1} 

\end{document}